\documentclass[preprint,prfluids,amsmath,amssymb]{revtex4-2}

\usepackage{dcolumn}
\usepackage{bm}
\usepackage[pdftex]{graphicx}
\usepackage{color}

\begin{document}

\title{Influence of different mutual friction models on two-way coupled quantized vortices and normal fluid in superfluid $^4$He}
\author{Hiromichi Kobayashi$^1$}\thanks{hkobayas@keio.jp}
\author{Satoshi Yui$^{2, 3}$}
\author{Makoto Tsubota$^{2, 3}$}
\affiliation{$^1$Department of Physics \& Research and Education Center for Natural Sciences, Hiyoshi Campus, Keio University, 4-1-1 Hiyoshi, Kohoku-ku, Yokohama 223-8521, Japan}\affiliation{$^2$Department of Physics, Osaka Metropolitan University, 3-3-138 Sugimoto, Sumiyoshi-ku, Osaka 558-8585, Japan}
\affiliation{$^3$Nambu Yoichiro Institute of Theoretical and Experimental Physics (NITEP), Osaka Metropolitan University, Osaka 558-8585, Japan}

\date{\today}

\begin{abstract}
We study the influence of two mutual friction models on quantized vortices and normal fluid using two-way coupled simulations of superfluid $^4$He.
The normal fluid is affected by quantized vortices via mutual friction.
A previous study [Y. Tang, {\it et al.} {Nat. Commun.} {\bf 14}, 2941 (2023)] compared the time evolutions of the vortex ring radius and determined that the self-consistent two-way coupled mutual friction (S2W) model yielded better agreement with the experimental results than the two-way coupled mutual friction (2W) model whose model parameters were determined through experiments with rotating superfluid helium.
In this study, we compare the two models in more detail in terms of the quantized vortex ring propagation, reconnection, and thermal counterflow.
We show that the S2W model exhibits better results than the 2W model on the microscopic scale near a quantized vortex, such as during quantized vortex ring propagation and reconnection, although the S2W model requires a higher spatial resolution.
For complex flows such as a thermal counterflow, the 2W model can be applied even to a low-resolution flow while maintaining the anisotropic normal fluid velocity fluctuations.
In contrast, the 2W model predicts lower normal fluid velocity fluctuations than the S2W model.
The two models show probability density functions with $- 3$ power-law tails for the normal fluid velocity fluctuations.
\end{abstract}

\maketitle

\newpage 

\section{Introduction}
\label{sec1}

At temperatures below 2.17 K, superfluid $^4$He is composed of an inviscid superfluid component and a viscous normal fluid component.
This is known as the two-fluid model \cite{Tisza,Landau}.
The circulation of the superfluid component is quantized and behaves as a quantized vortex line with a diameter of approximately 10$^{-8}$ cm.
In the hydrodynamics of superfluid $^4$He, the most extensively studied experimental phenomenon is the thermal counterflow \cite{TC-exp,TC-exp-review}.
In these experiments, two baths are filled with superfluid $^4$He and connected via a channel.
When one bath is heated, the normal fluid moves into the other bath through the channel.
However, the superfluid moves in the opposite direction through the channel to satisfy the conservation of mass.
As the heat flux increases, the relative velocity between the superfluid and normal fluid will also increase.
When the relative velocity exceeds a critical value, the quantized vortices become tangled \cite{Feynman}.
This phenomenon is known as quantum turbulence (QT).

A quantized vortex interacts with a normal fluid through mutual friction in superfluid $^4$He \cite{MF1,MF2,MF3,MF4}.
A model of mutual friction \cite{Hall-Vinen2,Schwarz-VFM-MF,Schwarz-VFM-wall} was proposed based on experimental data for uniformly rotating superfluid helium \cite{Hall-Vinen1,rho-mu-B}.
This model has been used for one-way coupled simulations, where the velocity profile of the normal fluid is prescribed, and quantized vortices in QT are affected via the mutual friction between the superfluid and normal fluid \cite{Schwarz-VFM-QT}.
Numerical simulations using the vortex filament model (VFM) with the full Biot-Savart law showed good agreement with the experimental results \cite{Full-BS,1wayQT-Baggaley,1wayQT-Lvov,1wayQT-Yui}.

The mutual friction model has also been used in two-way coupled simulations, where the normal fluid is locally affected by the quantized vortices via mutual friction.
The mutual friction model used in these two-way coupled simulations is referred to as the 2W model in this study.
The 2W model has been used in two-way coupled simulations with solid boundaries \cite{2way-channel,2way-duct}, producing the anomalous anisotropic velocity fluctuations of the normal fluid in counterflow experiments \cite{Yui-vf}.
However, the undisturbed velocity away from the core of the quantized vortex is assumed to be the normal fluid velocity in the 2W model \cite{Schwarz-VFM-MF}.

Two-way coupled simulations using mutual friction with a locally disturbed normal fluid velocity have also been conducted \cite{Science-ring,2way-Kivotides-energy,2way-Kivotides-spectrum}.
The concept of a self-consistent model \cite{Idoowu-self} was proposed and applied to simple configurations such as a quantized vortex ring \cite{Science-ring} and a straight quantized vortex \cite{Idoowu-self-straightline}.
The recent self-consistent model \cite{Kivotides-recent-self} was updated slightly to the present self-consistent two-way coupled mutual friction model \cite{Galantucci}, which is referred to as the S2W model in this study.
The S2W model adopts the theoretical friction force through a vortex line; consequently, no experimentally determined empirical parameters are required.

The time evolution of a single vortex ring radius obtained using the 2W and S2W models was compared with the experimental results, and the S2W model showed better agreement with the experimental results than the 2W model \cite{NC-ring}.
However, it is necessary to compare the performance of the two models for flows such as a vortex ring propagation, reconnection, and thermal counterflow.

Furthermore, monitoring the motion of solid hydrogen tracers in decaying QT has shown that the probability density function (PDF) of the superfluid velocity, $v$, is non-Gaussian with $1/v^3$ power-law tails owing to the motion of the quantized vortex \cite{m3tail}. These tails were reproduced in the PDF of the superfluid velocity obtained with a one-way coupled simulation of the VFM \cite{m3tail-Adachi} and the Gross-Pitaevskii equation in a turbulent atomic Bose-Einstein condensate \cite{m3tail-GP}. It would be interesting to observe the influence of the superfluid velocity on the PDF of the normal fluid using two-way coupled simulations with the two models.

In this study, two-way coupled simulations are conducted to examine the influence of the 2W and S2W models on the velocity fluctuations of the normal fluid.
The remainder of this paper is organized as follows:
In Section \ref{sec2}, the basic equations, mutual friction models, and numerical conditions are described.
In Section \ref{sec3}, we present and discuss the numerical results for the vortex ring propagation, reconnection, and thermal counterflow.
Finally, our conclusions are presented in Section \ref{sec4}.

\section{Basic equations and numerical conditions}
\label{sec2}

\subsection{Basic equations}

\begin{figure}
  \centering
  \includegraphics[width=0.8\linewidth]{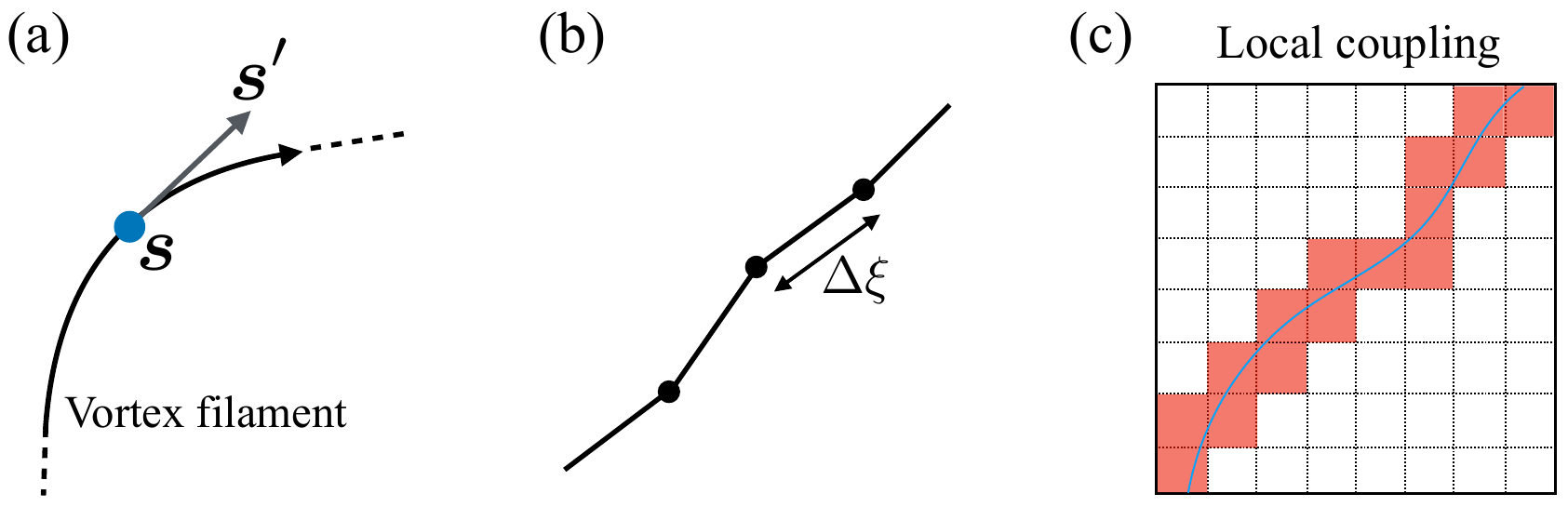}
  \caption{Schematic of a quantized vortex: (a) vortex filament and tangential vector $\bm{s}'$ at point $\bm{s}$, (b) separated segments with $\Delta \xi$, and (c) local coupling of a quantized vortex with the normal fluid discretized on grids. The quantized vortex is coupled with the normal fluid via mutual friction in the red mesh through which the quantized vortex passes.}
  \label{s-local}
\end{figure}

For the superfluid, the VFM \cite{Schwarz-VFM-MF,Schwarz-VFM-wall,Schwarz-VFM-QT} is used to describe the equation of motion of the quantized vortices.
Figure \ref{s-local} (a) presents a schematic of a vortex filament.
The tangential vector $\bm{s}'$ is defined as a unit vector along the vortex line at point $\bm{s}$.
The equation of motion for point $\bm{s}$ is as follows:
\begin{equation}
\frac{d \bm{s}}{dt}=\bm{v}_s+\alpha s'\times\bm{v}_{ns}-\alpha'\bm{s}'\times(\bm{s}'\times\bm{v}_{ns}),
\label{sdot}
\end{equation}
where $\bm{v}_s$ denotes the superfluid velocity, $\bm{v}_{ns}=\bm{v}_n-\bm{v}_s$, $\bm{v}_n$ is the normal fluid velocity, and $\alpha$ and $\alpha$' are the coefficients of mutual friction at a finite temperature for the 2W model \cite{rho-mu-B,Schwarz-VFM-MF,Schwarz-VFM-wall}.

The superfluid velocity at 0 K at position $\bm{r}$ is obtained from the induced velocity produced from segment $d\bm{s}_1$ at position $\bm{s}_1$ using the Biot-Savart law as follows:
\begin{equation}
\bm{v}_s(\bm{r})=\frac{\kappa}{4\pi}\int_\mathcal{L} \frac{(\bm{s}_1-\bm{r})\times d\bm{s}_1}{|\bm{s}_1-\bm{r}|^3}+\bm{v}_{s,b}+\bm{v}_{s,a},
\end{equation}
where $\kappa$ denotes the quantum circulation, $\mathcal{L}$ represents the integration along the vortex line, and 
$\bm{v}_{s,b}$ and $\bm{v}_{s,a}$ are the velocities induced from the boundaries and the uniform flow applied to the superfluid, respectively.

The momentum equations for the normal fluid are described by the Navier-Stokes equations:
\begin{equation}
\frac{\partial \bm{v}_n}{\partial t}+(\bm{v}_n\cdot\nabla)\bm{v}_n=-\frac{1}{\rho_n}\nabla P +\nu_n\nabla^2\bm{v}_n+\frac{1}{\rho_n}\bm{F}_{ns},
\label{NS}
\end{equation}
where the total density $\rho$ is composed of the superfluid density $\rho_s$ and normal fluid density $\rho_n$ as $\rho=\rho_s+\rho_n$; $P$ is the effective pressure, $\nu_n$ is the kinematic viscosity of the normal fluid, and $\bm{F}_{ns}$ denotes the mutual friction from the superfluid to the normal fluid.
As shown in Fig. \ref{s-local} (b), the mutual friction is calculated from each segment $\Delta \xi$ along the integral path $\mathcal{L'}$ in the normal fluid mesh $\Omega'(\bm{r})$ shown in Fig. \ref{s-local} (c) at position $\bm{r}$ as follows:
\begin{equation}
\bm{F}_{ns}=\frac{1}{\Omega'(\bm{r})}\int_{\mathcal{L'}(\bm{r})} \bm{f}(\xi)d\xi,
\label{Fns}
\end{equation}
where $\bm{f}(\xi)$ is the local mutual friction with arc length $\xi$.

\subsection{2W model}

In this study, we compare two models of mutual friction: the 2W model \cite{Schwarz-VFM-MF,Schwarz-VFM-wall} and the S2W model \cite{Galantucci}.

First, the 2W model \cite{Schwarz-VFM-MF,Schwarz-VFM-wall} is presented.
The mutual friction from the superfluid to the normal fluid $\bm{f}_{sn}$ is described based on experimental results \cite{Hall-Vinen1,rho-mu-B} as follows:
\begin{equation}
\bm{f}_{sn}=-\alpha \rho_s \kappa \bm{s}'\times \left( \bm{s}'\times \bm{v}_{ns} \right)-\alpha' \rho_s \kappa \bm{s}'\times \bm{v}_{ns}.
\label{fsn-2W}
\end{equation}
From $\bm{f}(\xi)=\bm{f}_{ns}=-\bm{f}_{sn}$, the local mutual friction in Eq. (\ref{Fns}) can be obtained.
\begin{equation}
\bm{f}(\xi)=\rho_s \kappa \left[ \alpha \bm{s}'\times(\bm{s}'\times\bm{v}_{ns})+\alpha' \bm{s}'\times\bm{v}_{ns} \right].
\end{equation}

The mutual friction $\bm{f}_{sn}$ can be interpreted from another perspective.
The vortex filament is affected by the Magnus force, $\bm{f}_M$.
\begin{equation}
\bm{f}_M=\rho_s \kappa \bm{s}'\times(\dot{\bm{s}}-\bm{v}_s),
\end{equation}
where $\dot{\bm{s}}=d\bm{s}/dt$.
In the 2W model, the drag force $\bm{f}_{D}$ is modeled using the drag coefficients $\gamma_0$ and $\gamma_0'$ \cite{rho-mu-B}.
\begin{equation}
\bm{f}_{D}=-\gamma_0 \bm{s}'\times \left[ \bm{s}'\times(\bm{v}_n-\dot{\bm{s}}) \right]+\gamma_0' \bm{s}'\times(\bm{v}_n-\dot{\bm{s}}).
\label{fD-2W}
\end{equation}
Note that $\gamma_0$ and $\gamma_0'$ correspond to $D$ and $D'$, respectively, in Ref. \cite{Schwarz-VFM-MF}.
In the 2W model, $\bm{f}_{sn}$ satisfies $\bm{f}_{sn}=\bm{f}_{D}$.
Because the inertia of the quantized vortex is negligible, the equation of motion can be expressed as follows:
\begin{equation}
\bm{f}_M+\bm{f}_{D}=\bm{0}.
\end{equation}
By eliminating $\dot{\bm{s}}$ using Eq. (\ref{sdot}), we obtain the following mutual friction coefficients:
\begin{equation}
\alpha = \frac{\rho_s \kappa \gamma_0}{\gamma_0^2+(\rho_s \kappa - \gamma_0')^2}, \quad 
\alpha' = \frac{\gamma_0^2-(\rho_s \kappa - \gamma_0')\gamma_0'}{\gamma_0^2+(\rho_s \kappa - \gamma_0')^2}.
\label{alpha}
\end{equation}

\subsection{S2W model}

This section discusses the S2W model \cite{Galantucci}.

In this model, the drag force is modeled using the drag coefficient $D$ as follows:
\begin{equation}
\bm{f}_D=-D \bm{s}'\times \left[ \bm{s}'\times(\bm{v}_n-\dot{\bm{s}}) \right].
\label{fD-S2W}
\end{equation}
When the Reynolds number Re$_{vortex}$ of the normal fluid based on the velocity induced by the vortex line is low ($10^{-5}\sim10^{-4}$), the coefficient of the drag force from the vortex line is analytically determined as follows \cite{drag-vortex-line}:
\begin{equation}
D=\frac{4 \pi \rho_n \nu_n}{\frac{1}{2}-\gamma-\ln \left( \mathrm{Re}_{vortex}\right)},
\end{equation}
\begin{equation}
\mathrm{Re}_{vortex}=\frac{|\bm{v}_{n\perp}-\dot{\bm{s}}|a_0}{4\nu_n},
\end{equation}
where the subscript $\perp$ denotes a component perpendicular to the vortex line,
$\gamma = 0.5772$ is the Euler-Mascheroni constant, and $a_0$ represents the vortex core size.
In this study, we set $a_0$ to $1.3\times10^{-10}$ m.

In this model, the Iordanskii force \cite {Iordanskii-1,Iordanskii-2} is taken into account:
\begin{equation}
\bm{f}_I=-\rho_n \kappa \bm{s}'\times(\bm{v}_n-\dot{\bm{s}}).
\label{fI-S2W}
\end{equation}

The equation of motion results in the following:
\begin{equation}
\bm{f}_M+\bm{f}_D+\bm{f}_I=\bm{0}.
\label{EOM-S2W}
\end{equation}

The mutual friction in this model is $\bm{f}_{sn}=\bm{f}_D+\bm{f}_I$.
Based on a comparison with Eq. (\ref{fsn-2W}), $\bm{f}_{sn}$ is modeled as follows:
\begin{equation}
\bm{f}_{sn}=\bm{f}_D+\bm{f}_I=-\beta \rho_s \kappa \bm{s}'\times \left( \bm{s}'\times \bm{v}_{ns} \right)-\beta' \rho_s \kappa \bm{s}'\times \bm{v}_{ns},
\end{equation}
where $\beta$ and $\beta'$ denote the coefficients of mutual friction in the S2W model.
Note that to ensure $\beta'$ is positive, the sign of $\beta'$ is opposite that in Ref. \cite{Galantucci,bundle-ring}.
The definition of $\beta'$ is consistent with $\alpha'$ in Eq. (\ref{sdot}) in the 2W model.
Comparing the two mutual friction models in Eq. (\ref{fD-2W}) and Eqs. (\ref{fD-S2W}) and (\ref{fI-S2W}), $\gamma_0$ and $\gamma_0'$ correspond to $D$ and $-\rho_n \kappa$, respectively.
Substituting these into Eq. (\ref{alpha}), the following mutual friction coefficients are obtained:
\begin{equation}
\beta = \frac{\rho_s \kappa D}{D^2+(\rho_s \kappa + \rho_n \kappa)^2}, \quad 
\beta' = \frac{D^2+(\rho_s \kappa + \rho_n \kappa)\rho_n \kappa}{D^2+(\rho_s \kappa + \rho_n \kappa)^2}.
\end{equation}
Finally, we obtain the equation of motion for $\bm{s}$ in the S2W model, similar to Eq. (\ref{sdot}):
\begin{equation}
\dot{\bm{s}}=\bm{v}_{s\perp}+\beta \bm{s}'\times \bm{v}_{ns}-\beta' \bm{s}'\times \left( \bm{s}'\times \bm{v}_{ns} \right).
\label{sdot-beta}
\end{equation}
A summary and comparison of the 2W and S2W models is presented in the Appendix.

Figure \ref{T-model} compares the coefficients of the 2W and S2W models as a function of the temperature, $T$.
The total coefficient of the S2W model is larger than that of the 2W model, and the ratio is stable at approximately two for temperatures higher than 1.6 K, as shown in Figs. \ref{T-model} (a) and \ref{T-model} (b).
The fractions of $\alpha$, $\beta$, $\alpha'$, and $\beta'$, i.e., $\alpha / \sqrt{\alpha^2 + \alpha'^2}$, $\beta / \sqrt{\beta^2 + \beta'^2}$, $\alpha' / \sqrt{\alpha^2 + \alpha'^2}$, and $\beta' / \sqrt{\beta^2 + \beta'^2}$, respectively, are shown in Figs. \ref{T-model} (c) and \ref{T-model }(d).
$\alpha$ is dominant at all temperatures in the 2W model, whereas $\beta$ decreases gradually with increasing temperature.
In contrast, $\beta'$ increases with increasing temperature.
These differences affect the strength of the mutual friction around the quantized vortex, as discussed in Section \ref{sec3}.

\begin{figure}
  \centering
  \includegraphics[width=1.0\linewidth]{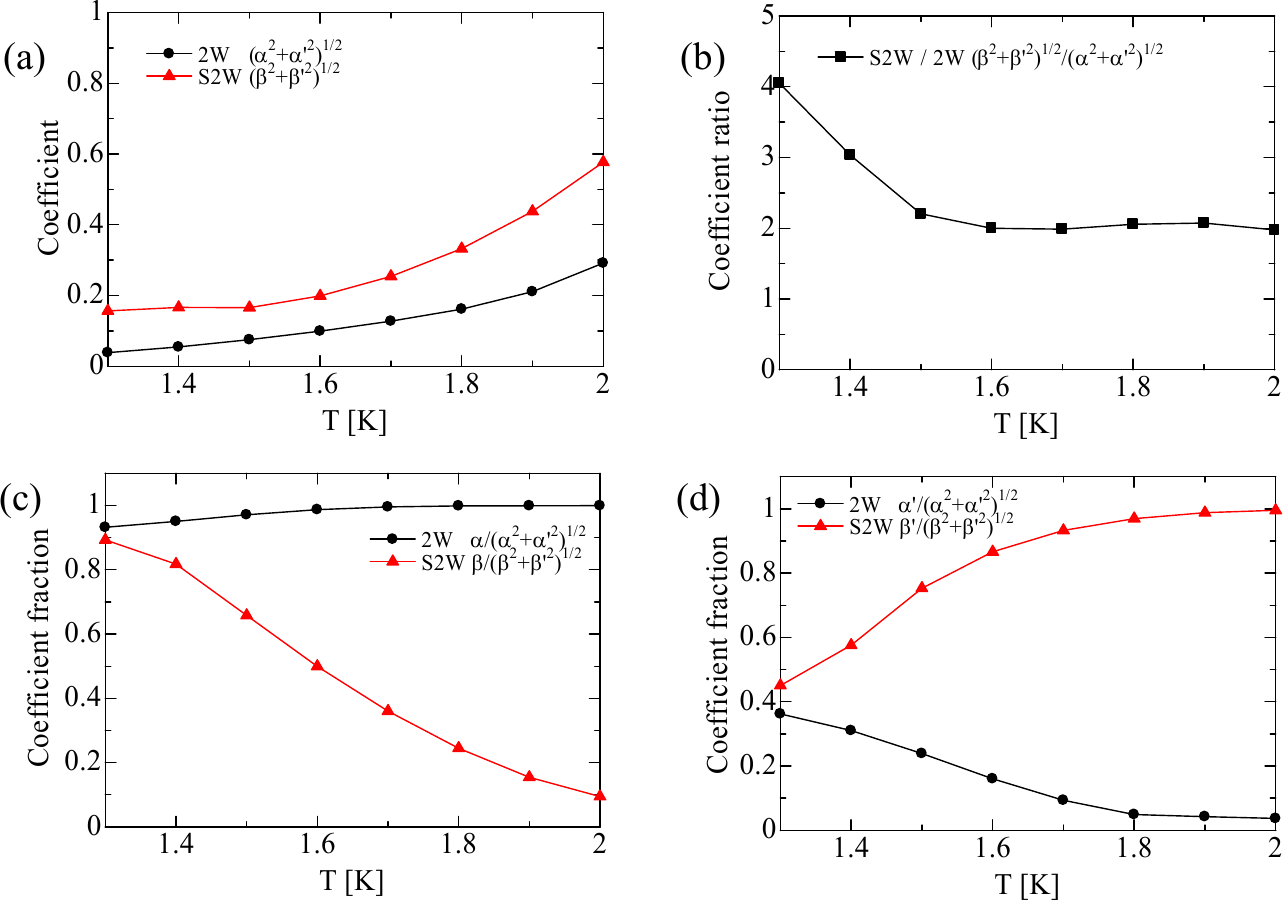}
  \caption{Comparison of the coefficients of the 2W and S2W models as a function of temperature, $T$: (a) total coefficient, (b) coefficient ratio, (c) coefficient fraction of $\alpha$ and $\beta$, and (d) coefficient fraction of $\alpha'$ and $\beta'$. Note that coefficient fraction (c) $+$ coefficient fraction (d) $\neq 1$.}
  \label{T-model}
\end{figure}

\subsection{Numerical methods and conditions}

\begin{figure}
  \centering
  \includegraphics[width=0.6\linewidth]{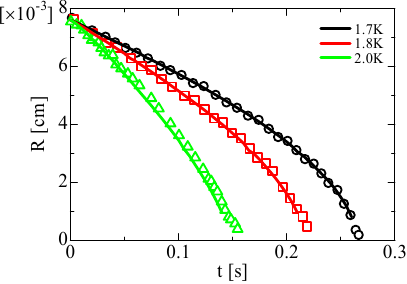}
  \caption{Time evolution of the radius of a quantized vortex ring in the S2W model for 1.7 K (black), 1.8 K (red) and 2.0 K (green); open symbols: reference results \cite{bundle-ring}.}
  \label{ring-radius-beta}
\end{figure}

Time integration of Eqs. (\ref{sdot}) or (\ref{sdot-beta}) was performed using the fourth-order accuracy Runge-Kutta method with $\Delta t=$ 0.0001 s.
The spatial resolution of $\xi$ between discrete points was set to 0.0008 cm $< \Delta \xi <$ 0.0024 cm.
Two filaments were considered to be reconnected if they approached within $\xi_{min}=$ 0.0008 cm \cite{Full-BS}.
Short filaments of less than $5\Delta \xi_{min}$ were removed \cite{removal-filament}.

Equations (\ref{NS}) for the normal fluid were discretized using the second-order accuracy finite difference method.
The simplified Maker and Cell method \cite{MAC} was used to couple the velocity and pressure, and the fast Fourier transform was used to solve the Poisson equation for the pressure.

A temperature, $T$, of 1.9 \rm{K} was considered.
A box size of $D_x=D_y=D_z=$ 1 mm was used, and the number of grid points for the normal fluid was set to $N_x=N_y=N_z=120$.

Periodic conditions were adopted, and the uniform flow $\bm{v}_{s,a}=-\rho_n \bm{V}_n /\rho_s$ based on the mass conservation law was applied to the thermal counterflow, where $V_n$ is the mean velocity of the normal fluid.
In this study, $V_n$ of 2.5 mm/s and 5.0 mm/s in the $x$ direction were considered.

For the quantized vortex ring propagation, the initial radius was set to 0.02 cm.
For the reconnection, two quantized vortices cross at a 90-degree angle, and the initial distance was set to 0.002 cm.

Our implementation of the S2W model was then validated.
Figure \ref{ring-radius-beta} shows the time evolution of the radius of a quantized vortex ring using the S2W model for 1.7, 1.8, and 2.0 K.
The initial radius was set to $7.6 \times 10^{-3}$ cm.
The results showed good agreement with the data presented in Ref. \cite{bundle-ring}.

\section{Numerical results and discussion}
\label{sec3}

\subsection{Quantized vortex ring propagation}

\begin{figure}
  \centering
  \includegraphics[width=1.0\linewidth]{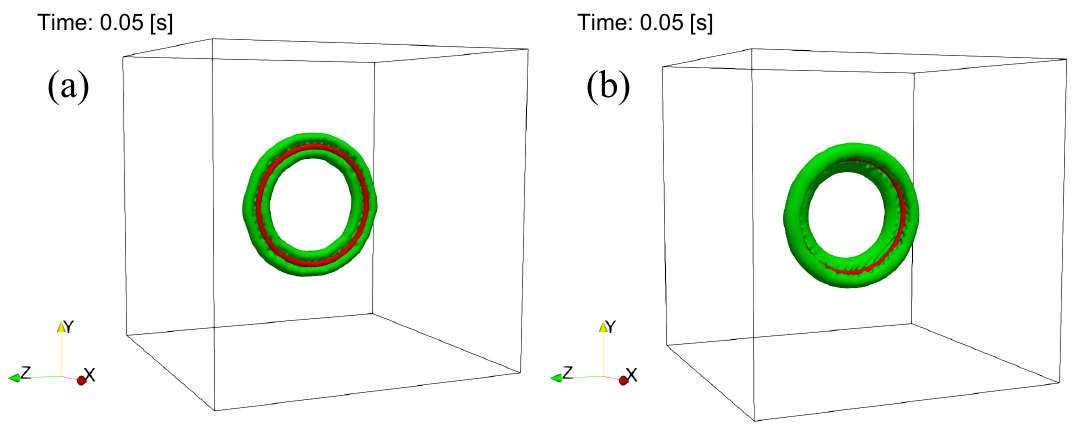}
  \caption{Quantized vortex ring propagation at 1.9 K and 0.05 \rm{s} in the (a) 2W model and (b) S2W model; red color: quantized vortex, green color: normal fluid vortex tube ($Q=0.5$ $s^{-2}$).}
  \label{ring}
\end{figure}

Figure \ref{ring} shows the quantized vortex ring propagation at 0.05 \rm{s} for the 2W and S2W models; the quantized vortex is visualized in red and the normal fluid vortex tube is displayed in green using the second invariant of the velocity gradient tensor $Q = 0.5$ $s^{-2}$ \cite{Q-value}.
The second invariant is defined as $Q = (W_{ij} W_{ij} - S_{ij} S_{ij})/2$ using the velocity strain tensor $S_{ij}=(\partial v_{n,j}/\partial x_i + \partial v_{n,i}/\partial x_j)/2$ and the vorticity tensor $W_{ij}=(\partial v_{n,j}/\partial x_i - \partial v_{n,i}/\partial x_j)/2$.
For the 2W model, a pair of rings of normal fluid vortex tubes are located inside and outside the quantized vortex ring in the radial direction.
In the S2W model, the inner vortex tube remains slightly behind the quantized vortex ring, and the outer vortex tube propagates slightly ahead of the quantized vortex ring.
This difference between the models has also been observed at 1.65 \rm{K} \cite{NC-ring}.
This is due to the coefficients of mutual friction of $\alpha$ ($\alpha'$) and $\beta$ ($\beta'$).

\begin{figure}
  \centering
  \includegraphics[width=0.4\linewidth]{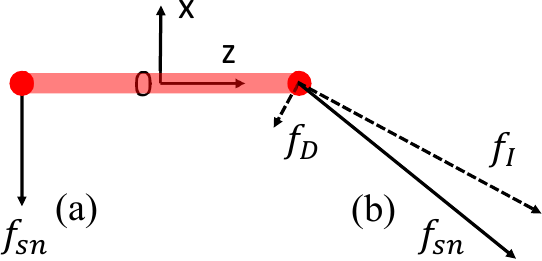}
  \caption{Mutual friction to the vortex ring $f_{sn}$ on the $x-z$ plane at 1.9 K and 0.05 \rm{s} in the (a) 2W model $\bm{f}_{sn} = (-8.62\times10^{-10}, 9.50\times10^{-12})$ N/m and (b) S2W model with $\bm{f}_D = (-3.12\times10^{-10}, -1.69\times10^{-10})$ N/m and $\bm{f}_I = (-9.25\times10^{-10}, 1.71\times10^{-9})$ N/m where $\bm{f}_{sn}=\bm{f}_D+\bm{f}_I=(-1.24\times10^{-9}, 1.54\times10^{-9})$ N/m; red color: quantized vortex; O indicates the center of the quantized vortex ring.}
  \label{fsn}
\end{figure}

The mutual friction to the vortex ring $f_{sn}$ on the $x-z$ plane at 0.05 \rm{s} is shown in Fig. \ref{fsn}.
The mutual friction in the 2W model acts opposite to the propagation direction of the ring.
However, the mutual friction in the S2W model acts in a diagonal direction.
These orientations of the mutual friction are consistent with those shown in Figs. \ref{T-model} (c) and \ref{T-model} (d).
As the temperature increases, the direction of the mutual friction in the S2W model rotates from $-x$ to $z$.
The mutual friction in the S2W model is approximately 2.3 times stronger than that in the 2W model.
This result is consistent with the coefficient ratio in Fig. \ref{T-model}(b).
$f_D$ is much weaker than $f_I$ and acts inside the ring, whereas $f_I$ acts outside the ring, where $\bm{f}_{sn}=\bm{f}_D+\bm{f}_I$. Substituting Eq. (\ref{sdot-beta}) into Eqs. (\ref{fD-S2W}) and (\ref{fI-S2W}) results in the following:
\begin{equation}
\bm{f}_D=-D \left[ (1-\beta') \bm{s}'\times (\bm{s}'\times \bm{v}_{ns})+\beta \bm{s}'\times \bm{v}_{ns} \right] ,
\end{equation}
\begin{equation}
\bm{f}_I=\rho_n \kappa \left[ \beta \bm{s}'\times (\bm{s}'\times \bm{v}_{ns})-(1-\beta') \bm{s}'\times \bm{v}_{ns} \right],
\end{equation}
where
\begin{equation}
D=\frac{\rho_s \kappa \beta}{\beta^2+(1-\beta')^2},
\end{equation}
\begin{equation}
\rho_n \kappa =\frac{\rho_s \kappa \left[-\beta^2+\beta'(1-\beta')\right]}{\beta^2+(1-\beta')^2}.
\end{equation}

It is worth noting that the mutual friction predicted by the S2W model should approach that predicted by the 2W model at the coarse-graining limit.
If experiments are performed to investigate the location of normal fluid vortex tubes around a vortex ring, the accuracy of the S2W model will be improved.

\begin{figure}
  \centering
  \includegraphics[width=0.6\linewidth]{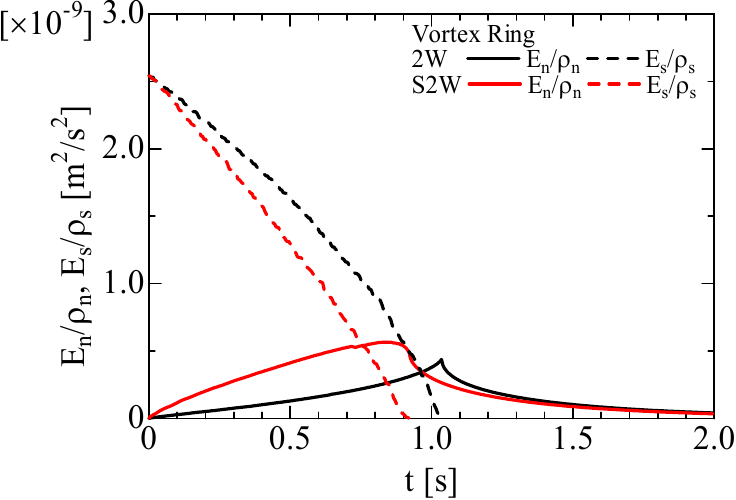}
  \caption{Time evolution of the kinetic energy per unit density of the normal fluid (solid line) and superfluid (dashed line) at 1.9 K during vortex ring propagation; black color: 2W model, red color: S2W model.}
  \label{t-ring}
\end{figure}

The time evolution of the kinetic energy per unit density of the normal fluid and superfluid is shown in Fig. \ref{t-ring}.
The kinetic energies of the normal fluid and superfluid are defined as follows:
\begin{equation}
\frac{E_n}{\rho_n} = \frac{1}{2}v_n^2, \quad \frac{E_s}{\rho_s} = \frac{1}{2}v_s^2.
\end{equation}
The energy of the superfluid is gradually decreased by transferring energy to the normal fluid via mutual friction.
The superfluid vortex ring in the 2W model disappears at 1.0 s, whereas that in the S2W model annihilates at 0.9 s.
The superfluid energy in the 2W model maintains a longer lifetime than that in the S2W model.
This result is consistent with that at 1.65 K \cite{NC-ring}.
The normal fluid energy in the S2W model increases faster than that in the 2W model.
After annihilation of the superfluid vortex ring, the normal fluid energy decreases owing to the viscosity of the normal fluid.

\begin{figure}
  \centering
  \includegraphics[width=1.0\linewidth]{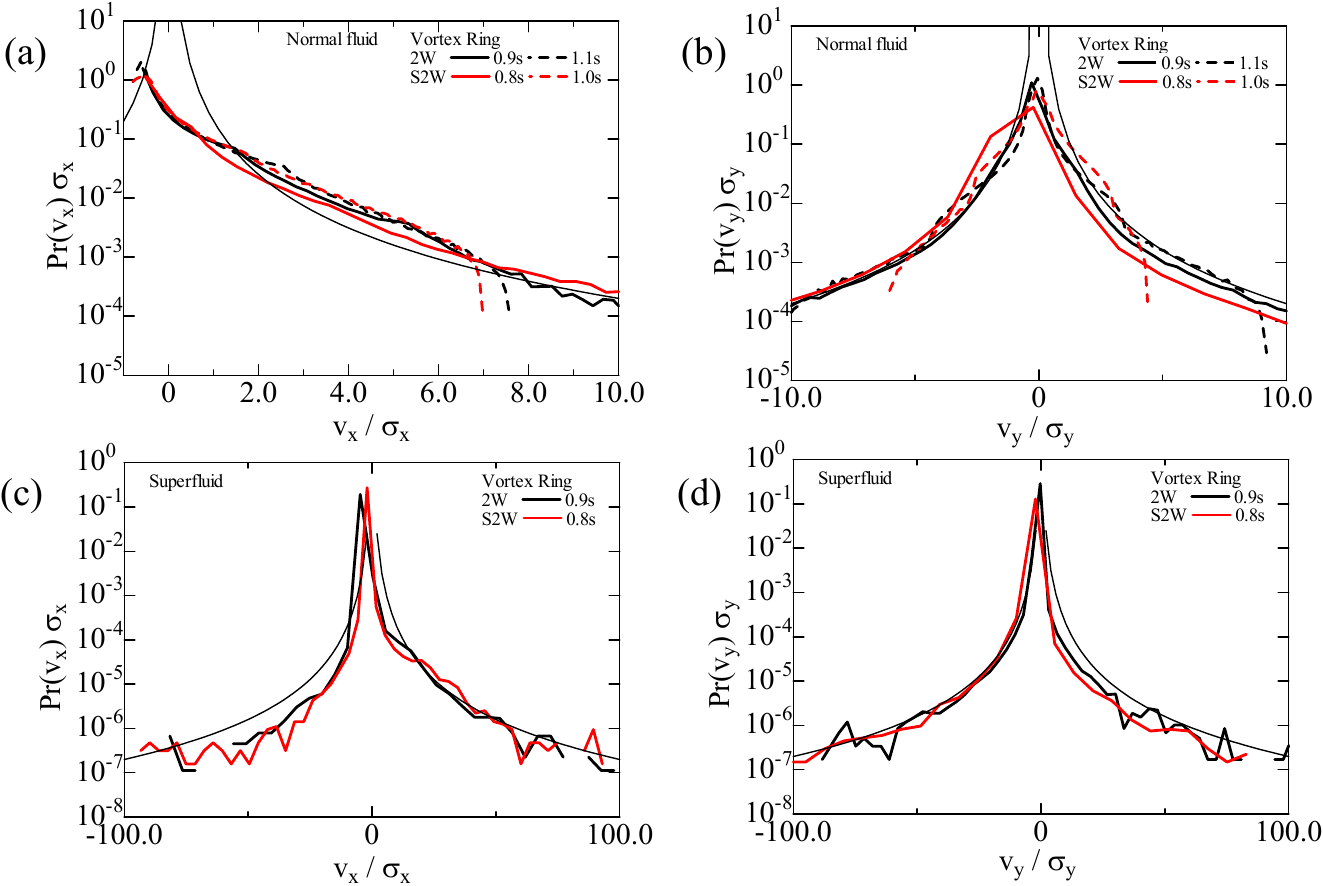}
  \caption{PDFs of velocity fluctuations at 1.9 K during vortex ring propagation for the normal fluid in the (a) $x$ (propagation) and (b) $y$ (radial) directions and the superfluid in the (c) $x$ and (d) $y$ directions; black color: 2W model at 0.9 s (solid line) and 1.1 s (dashed line), red color: S2W model at 0.8 s (solid line) and 1.0 s (dashed line); solid line: immediately before annihilation, dashed line: immediately after annihilation; fine solid line: $1/v^3$.}
  \label{pdf-ring}
\end{figure}

Next, the PDFs are compared.
Figure \ref{pdf-ring} shows the PDFs of the velocity fluctuations during the vortex ring propagation.
It is known that superfluid velocity fluctuations exhibit strong non-Gaussian PDFs with $- 3$ power-law tails \cite{m3tail}, as shown in Figs. \ref{pdf-ring} (c) and \ref{pdf-ring} (d).
The superfluid PDFs shown with solid lines correspond to the event immediately before annihilation of the quantized vortex rings.
No dashed lines are shown because the quantized vortex ring has disappeared.

Normal fluid velocity fluctuations are known to have Gaussian PDFs, although those velocity gradients yield non-Gaussian PDFs \cite{exp-Gauss,DNS-Gauss,DNS-Gauss2}.
The PDFs of the normal fluid velocity fluctuations in the $y$ (radial) direction have $- 3$ power-law tails, as shown by the fine solid line.
The PDFs are affected by the superfluid fluctuations via mutual friction.
The dashed lines indicate the normal fluid PDFs immediately after the annihilation of the quantized vortex rings.
The PDFs with strong fluctuations are reduced because the superfluid vortex rings disappear.
Figure \ref{pdf-ring} (a) presents the normal fluid PDFs in the $x$ (propagation) direction.
The PDFs have $- 3$ power-law tails, which are affected by the superfluid fluctuations via mutual friction.
However, the tails exist only in the propagation direction, i.e., in the positive $x$ direction.
The peaks of the PDFs are located at approximately $- 0.5$.
This is due to the normal fluid backflow in the inner region of the quantized vortex ring.
As shown in Fig. \ref{ring}, the inner normal fluid vortex caused by the quantized vortex ring rotates through mutual friction and produces the backflow in the inside of the vortex ring.
The backflow is rectified by the normal fluid contraction flows induced by mutual friction.
Consequently, it is believed that almost no negative fluctuations occur.
In terms of the PDFs, almost no difference between the models is observed.

\subsection{Reconnection of two quantized vortices}

\begin{figure}
  \centering
  \includegraphics[width=1.0\linewidth]{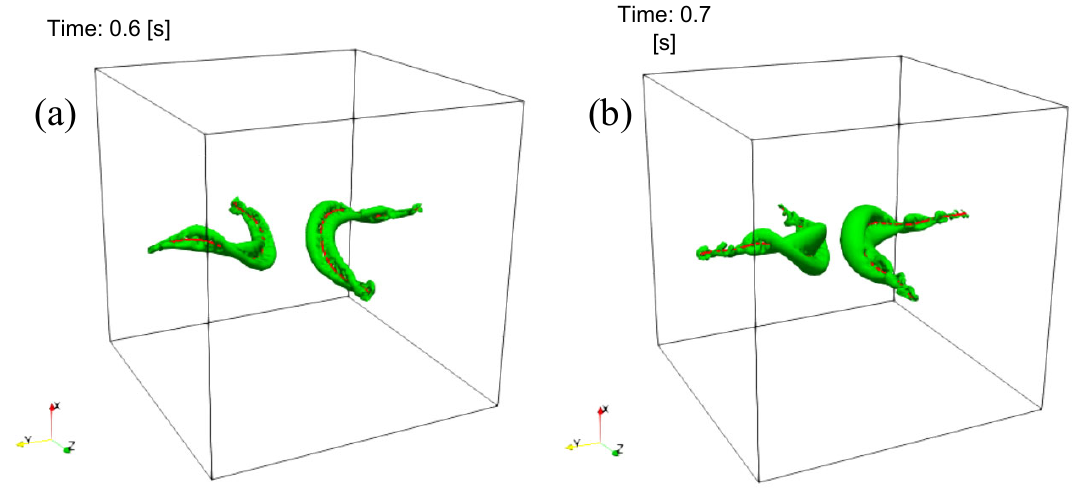}
  \caption{Vortex tubes and quantized vortices after reconnection in the (a) 2W model and (b) S2W model; red color: quantized vortex, green color: normal fluid vortex tube ($Q=0.5$ $s^{-2}$).}
  \label{rec}
\end{figure}

\begin{figure}
  \centering
  \includegraphics[width=0.6\linewidth]{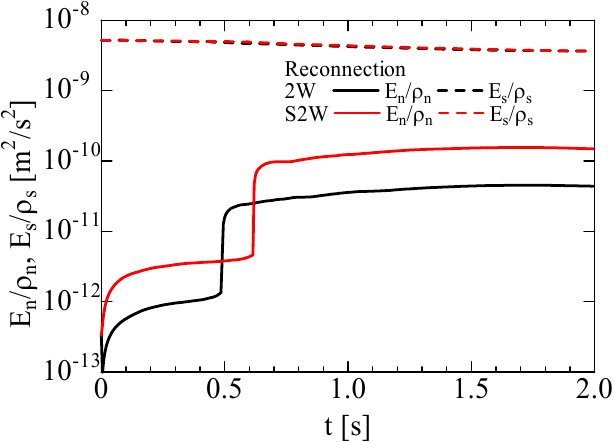}
  \caption{Time evolution of the kinetic energy per unit density of the normal fluid (solid line) and superfluid (dashed line) during reconnection; black color: 2W model, red color: S2W model.}
  \label{t-rec}
\end{figure}

\begin{figure}
  \centering
  \includegraphics[width=1.0\linewidth]{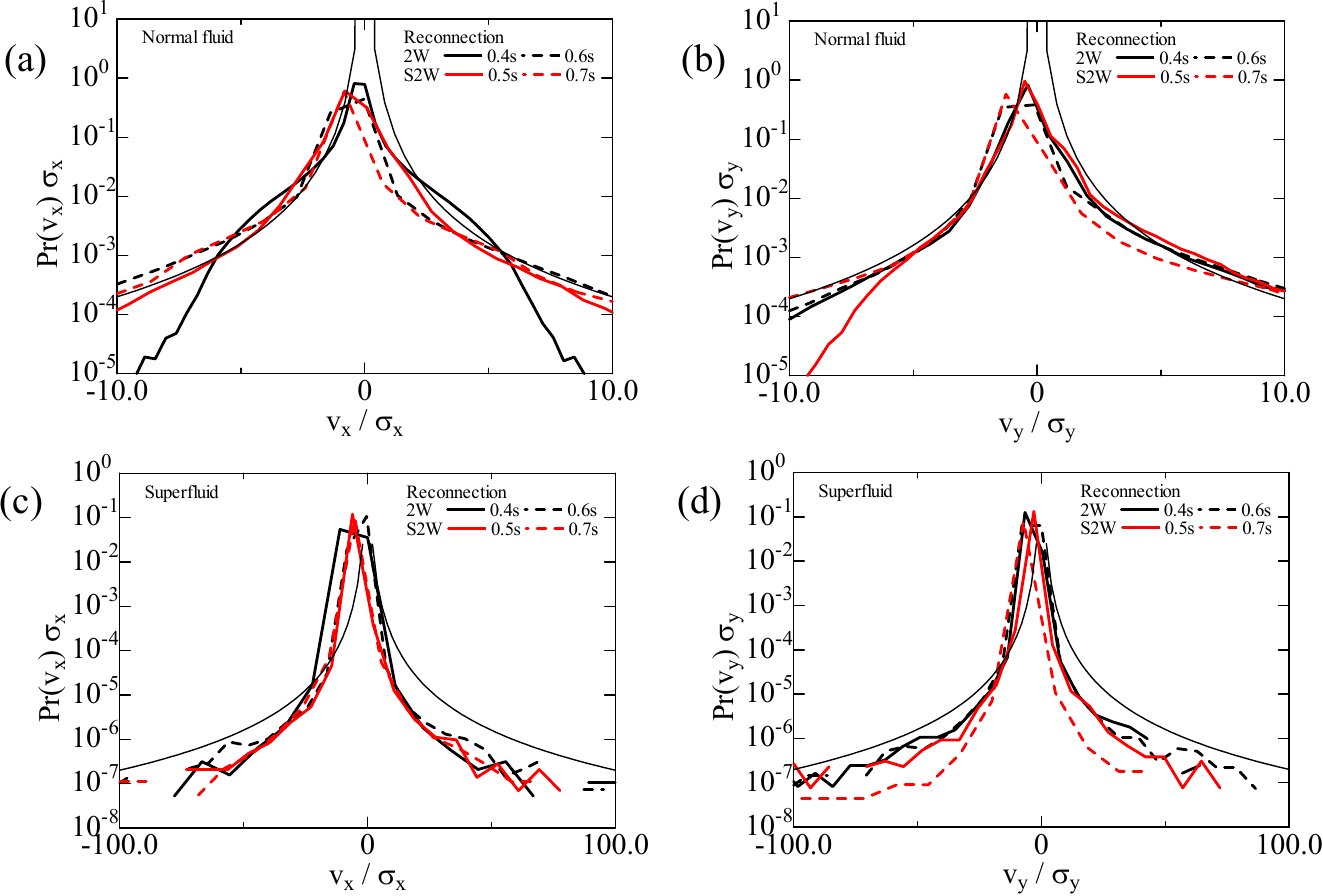}
  \caption{PDFs of the velocity fluctuations during reconnection for the normal fluid in the (a) $x$ (quantized vortex propagation) and (b) $y$ (perpendicular) directions and the superfluid in the (c) $x$ and (d) $y$ directions; black color: 2W model at 0.4 s (solid line) and 0.6 s (dashed line), red color: S2W model at 0.5 s (solid line) and 0.7 s (dashed line); solid line: immediately before reconnection, dashed line: immediately after reconnection; fine solid line: $1/v^3$.}
  \label{pdf-rec}
\end{figure}

\begin{figure}
  \centering
  \includegraphics[width=1.0\linewidth]{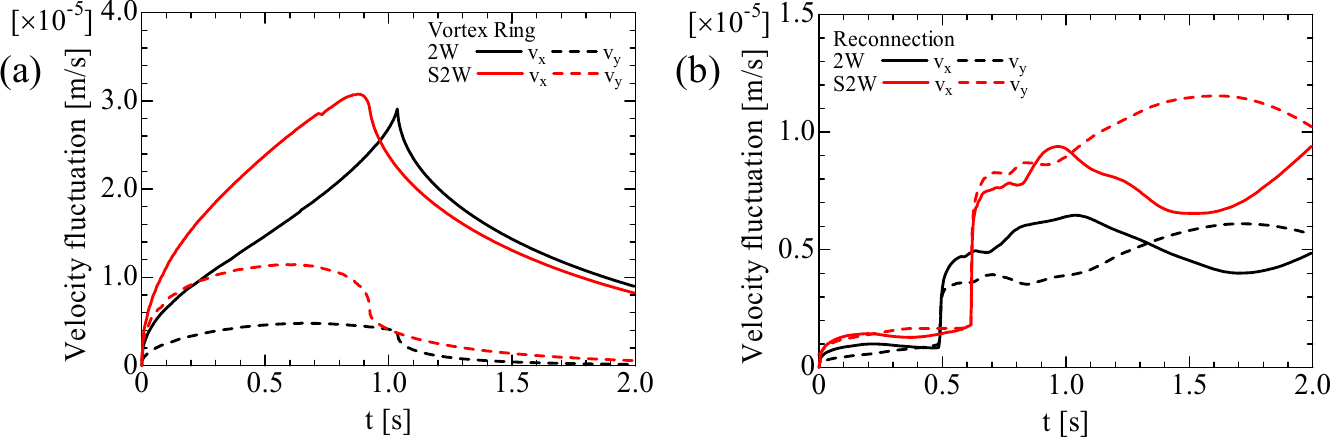}
  \caption{Velocity fluctuations of the normal fluid in the $x$ (solid line) and $y$ (dashed line) directions during (a) vortex ring propagation and (b) reconnection; black color: 2W model, red color: S2W model.}
  \label{t-xy-ring-rec}
\end{figure}

The vortex tubes and quantized vortices after reconnection are shown in Fig. \ref{rec}; the quantized vortex is shown in red, and the normal fluid vortex tube is depicted in green using the second invariant $Q = 0.5$ s$^{-2}$ of the velocity gradient tensor.
For the S2W model, the spiral vortex tubes that emerge around the quantized vortex line have a stronger twist than those in the 2W model.

Figure \ref{t-rec} shows the time evolution of the kinetic energy per unit density of the normal fluid and superfluid during reconnection.
Before reconnection, the two quantized vortices are twisted while approaching each other owing to the induced velocity from the other quantized vortex.
The motion of the quantized vortices produces a weak vortex tube of normal fluid via mutual friction.
Consequently, the energy of the superfluid is transferred to the normal fluid.
An abrupt energy transfer from the superfluid to the normal fluid occurs at approximately 0.5 s.
Because the S2W model has stronger mutual friction than the 2W model, the reconnection is delayed.
The normal fluid in the S2W model has a higher energy than that in the 2W model owing to the stronger mutual friction.

Figure \ref{pdf-rec} shows the PDFs of the velocity fluctuations during reconnection.
The PDFs of the normal fluid and superfluid exhibit $- 3$ power-law tails in all directions.
This is due to the superfluid fluctuations.
Although there is almost no difference in the superfluid PDFs between the two models, the normal fluid PDFs drawn with solid lines appear different, as shown in Figs. \ref{pdf-rec} (a) and \ref{pdf-rec} (b).
The solid lines correspond to the time immediately before reconnection.
The normal fluid fluctuations remain weak, as shown in Fig. \ref{t-rec} because the influence of the mutual friction on the normal fluid is weak.
Consequently, the PDFs do not yet exhibit $- 3$ power-law long tails, i.e., the PDFs show the results during a transient process.
Nevertheless, the two models produce different PDFs.
This is due to the difference in the location of the normal fluid vortex tubes around the quantized vortex, as shown in Fig. \ref{ring}.

Figure \ref{t-xy-ring-rec} (a) shows the velocity fluctuations of the normal fluid in the $x$ and $y$ directions during vortex ring propagation.
The velocity fluctuation is defined as the root-mean-square velocity $\sqrt{v_{n,i}^2}$ $(i=x, y)$.
The normal fluid in the S2W model receives a greater amount of energy than that in the 2W model.
The velocity fluctuation in the $x$ direction is stronger than that in the $y$ direction.
This is a result of jet formation of the normal fluid due to the local mutual friction $\bm{f}_{ns}=-\bm{f}_{sn}$ as shown in Fig. \ref{fsn}.
Figure \ref{t-xy-ring-rec} (b) shows the velocity fluctuations of the normal fluid in the $x$ and $y$ directions during reconnection.
As an initial condition, two straight quantized vortex lines crossed at a 90-degree angle are set at a certain distance in the $x$ (vertical) direction.
Immediately after reconnection, the 2W model yields a stronger fluctuation in the $x$ direction, whereas the S2W model yields a stronger fluctuation in the $y$ direction.
This is due to the location of the normal fluid vortex tubes produced around the quantized vortex line, as shown in Figs. \ref{ring} and \ref{rec}.

\subsection{Thermal counterflow}

\begin{figure}
  \centering
  \includegraphics[width=0.8\linewidth]{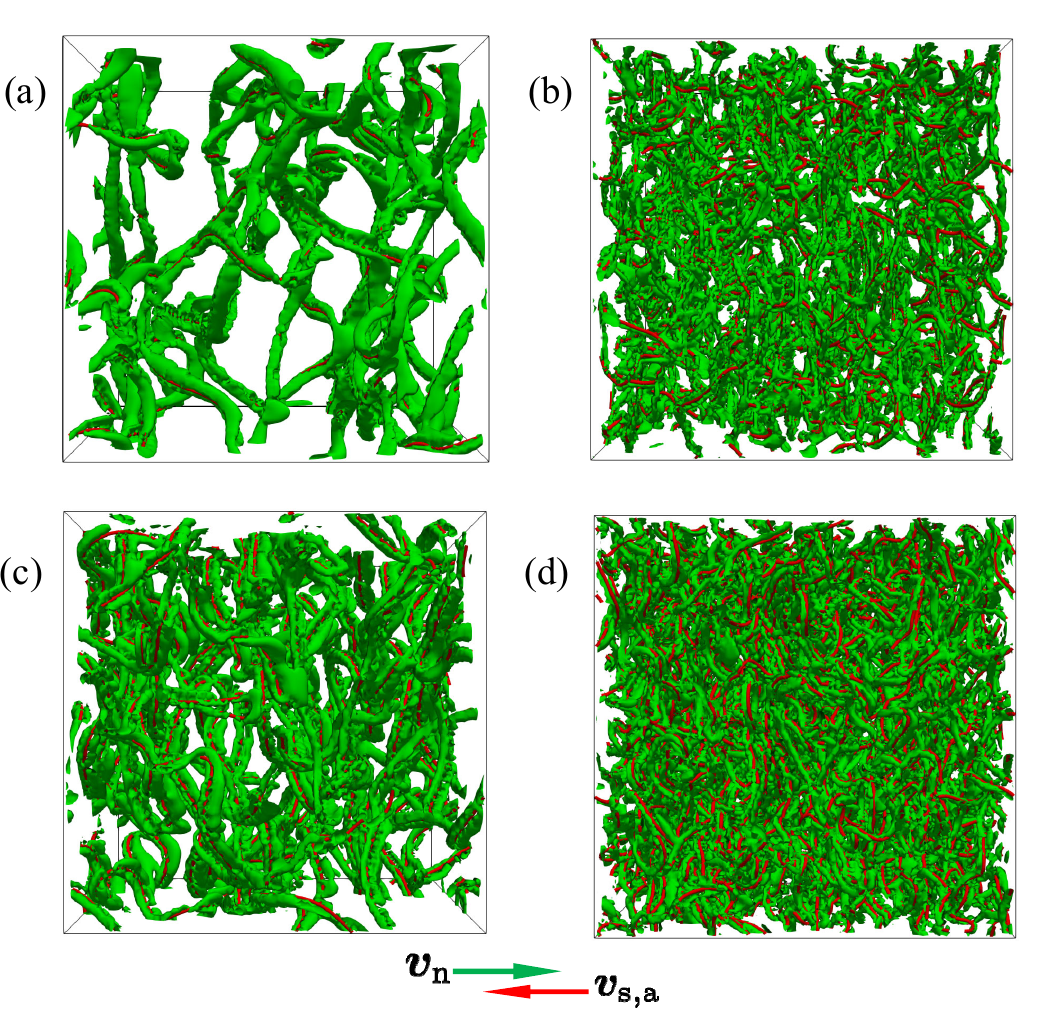}
  \caption{Quantized vortices (red) and normal fluid vortex tubes (green) in a thermal counterflow for the 2W model (a) $V_n = 2.5$ mm/s ($Q=10$ $s^{-2}$), (b) $V_n = 5.0$ mm/s ($Q=100$ $s^{-2}$) and S2W model (c) $V_n = 2.5$ mm/s ($Q=100$ $s^{-2}$), (d) $V_n = 5.0$ mm/s ($Q=1000$ $s^{-2}$); $V_n$ denotes the mean normal fluid velocity.}
  \label{cf}
\end{figure}

\begin{figure}
  \centering
  \includegraphics[width=0.6\linewidth]{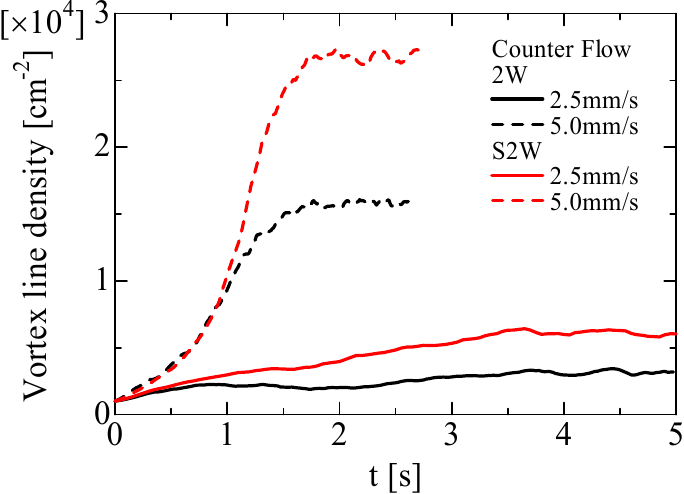}
  \caption{Time evolution of the vortex line density in the thermal counterflow for the 2W model and S2W model at $V_n = 2.5$ and 5.0 mm/s.}
  \label{vld}
\end{figure}

\begin{figure}
  \centering
  \includegraphics[width=0.6\linewidth]{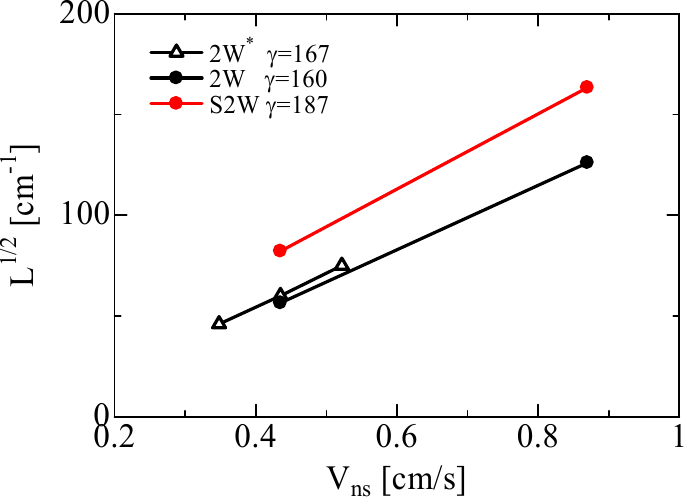}
  \caption{
Average vortex line density in the statistically steady state of Fig. \ref{vld} as a function of the mean relative velocity, $V_{ns}$, where $\gamma = L^{1/2}/(V_{ns}-v_0)$ and $v_0$ is a small, adjustable parameter on the order of 1 cm s$^{-1}$; 2W$^*$ denotes the result in our previous study using the 2W model \cite{Yui-vf}.
}
  \label{gamma}
\end{figure}

This section considers the thermal counterflow.
Figure \ref{cf} shows a snapshot of the quantized vortices and normal fluid vortex tubes in a thermal counterflow.
The vortex tubes are produced by mutual friction with quantized vortices.
The S2W model yields a high density of vortex lines and strong vortex tubes.
The vortex line density is defined as the vortex line length per unit volume.
The time evolution of the vortex line density in the thermal counterflow is shown in Fig. \ref{vld}.
The density gradually increases at 2.5 mm/s, whereas it increases rapidly at 5.0 mm/s.
Subsequently, the density reaches a statistically steady state.
The S2W model yields a density that is approximately twice that of the 2W model.
Figure \ref{gamma} shows the average vortex line density in the statistically steady state of Fig. \ref{vld} as a function of the mean relative velocity, $V_{ns}$.
The slope parameter $\gamma = L^{1/2}/(V_{ns}-v_0)$ is presented in the figure, where $v_0$ is a small, adjustable parameter on the order of 1 cm s$^{-1}$.
The value $\gamma = 167$ of 2W$^*$ is the reference value in our previous study using the 2W model \cite{Yui-vf} and is consistent with the present result of $\gamma = 160$ obtained using the 2W model.
The S2W model yields a higher $\gamma = 187$ than the 2W model, although the experimental value is $\gamma \sim 130$ s/cm$^2$ \cite{gamma1_1,gamma1_2}.

The Vinen equation \cite{TC-exp-review,MF3} can be expressed as follows:
\begin{equation}
\frac{dL}{dt} = A |V_{ns}| L^{3/2} - B L^2,
\end{equation}
where $A$ and $B$ denote the coefficients of the generation and decay rates, respectively.
In the steady state, $L^{1/2} = A/B |V_{ns}|$ is obtained.
Here, $A/B \sim \gamma$.
As shown in Fig. \ref{t-rec}, the stronger mutual friction in the S2W model leads to a slower approach between two quantized vortices.
Consequently, the decay rate becomes low, and thus the S2W model estimates a larger $\gamma$ than the 2W model.


\begin{figure}
  \centering
  \includegraphics[width=1.0\linewidth]{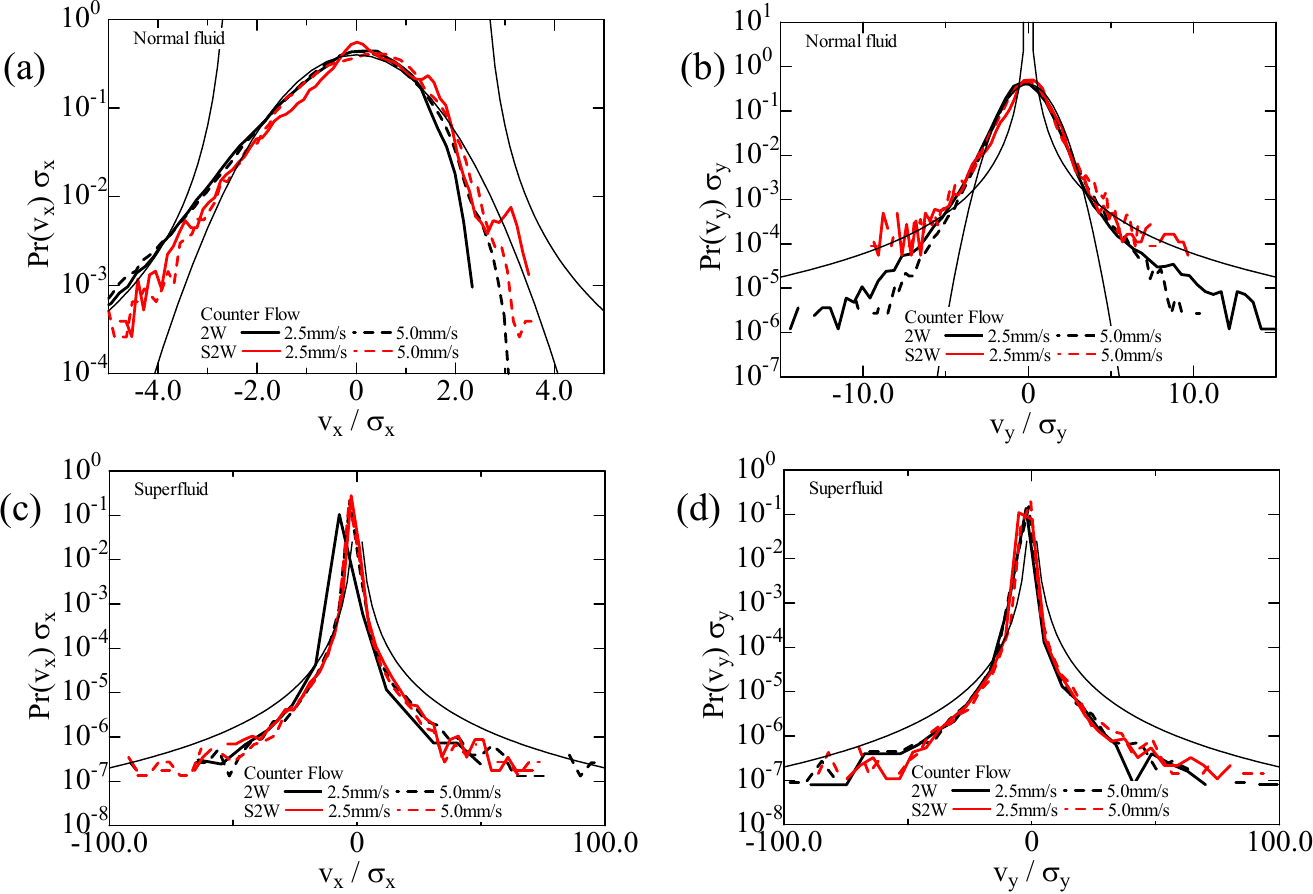}
  \caption{PDFs of velocity fluctuations in the thermal counterflow for the normal fluid in the (a) $x$ (streamwise) and (b) $y$ (perpendicular) directions and the superfluid in the (c) $x$ and (d) $y$ directions; black color: 2W model, red color: S2W model; solid line: mean normal fluid velocity of 2.5 mm/s, dashed line: 5.0 mm/s; fine solid line: $1/v^3$ and Gaussian profile for (a) and (b).}
  \label{pdf-cf}
\end{figure}

The PDFs of the velocity fluctuations in the thermal counterflow in the statistically steady state are shown in Fig. \ref{pdf-cf}.
Two mean normal fluid velocities of 2.5 mm/s and 5.0 mm/s are examined.
There is no difference in the direction of fluctuations in the PDFs of the superfluid.
The PDFs of the normal fluid in the $x$ and $y$ directions exhibit Gaussian distributions with weak fluctuations.
As shown in Fig. \ref{pdf-cf} (b), strong fluctuations exhibit $- 3$ power-law tails.
The S2W model yields stronger fluctuations than the 2W model owing to the stronger mutual friction.
The mean normal fluid velocity has a weak influence on the intensity of the PDFs of the normal fluid.
The PDFs in the $x$ (streamwise) direction appear asymmetric in Fig. \ref{pdf-cf} (a).
The positive fluctuations exhibit sub-Gaussian distributions, whereas the negative fluctuations produce $- 3$ power-law tails.
In this study, the normal fluid flows in the positive $x$ direction, and the superfluid moves in the negative $x$ direction during the counterflow.
If quantized vortex rings are generated, they tend to propagate in the negative $x$ direction.
Consequently, long-tail PDFs are produced in the propagation direction of the quantized vortex rings, as shown in Fig. \ref{pdf-ring} (a).

\begin{figure}
  \centering
  \includegraphics[width=1.0\linewidth]{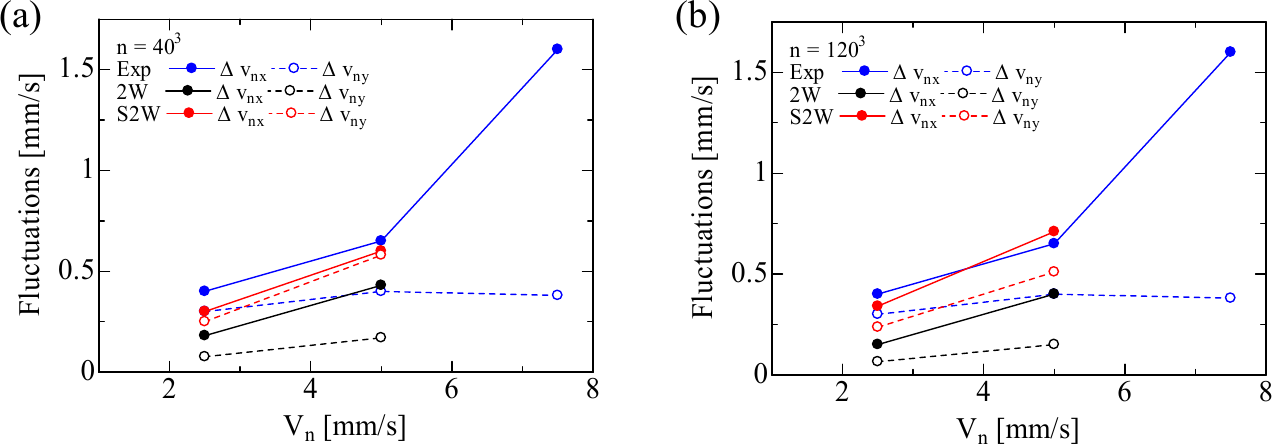}
  \caption{Normal fluid velocity fluctuations in the $x$ (streamwise, solid line) and $y$ (perpendicular, dashed line) directions as a function of the mean normal fluid velocity, $V_n$, in the thermal counterflow for grid numbers of (a) $40^3$ and (b) $120^3$; Exp: experimental results (blue) \cite{avf}, 2W model (black), and S2W model (red).}
  \label{vn-vf}
\end{figure}

Figure \ref{vn-vf} shows the normal fluid velocity fluctuations in the 2W and S2W models as a function of the mean normal fluid velocity in the thermal counterflow.
At a low resolution, the 2W model reproduces the anisotropic fluctuations, whereas the S2W model produces fewer anisotropic fluctuations.
At high resolution, the 2W model yields almost the same fluctuations as the low-resolution model, whereas the S2W model generates anisotropic fluctuations.
The S2W model requires higher resolution and yields higher fluctuations than the 2W model.
However, the S2W model predicts the normal fluid fluctuations better than the 2W model.

\section{Conclusions}
\label{sec4}
We investigated the performance of two different mutual friction models, i.e., the 2W and S2W models, on quantized vortices and normal fluid using two-way coupled simulations of superfluid $^4$He.
In the quantized vortex ring propagation and reconnection, the normal fluid vortex tube induced by mutual friction is produced at slightly different locations around the quantized vortex in each model.
The normal fluid velocity fluctuations in the S2W model are stronger than those in the 2W model, whereas the probability density functions produced by the two models show negligible differences.

The S2W model is better suited for describing the normal fluid physics on a microscopic scale near a quantized vortex, such as during quantized vortex ring propagation and reconnection.
For complex flows, such as a thermal counterflow, the 2W model can represent a low-resolution flow while maintaining anisotropic fluctuations.
However, the S2W model requires higher resolution and yields higher fluctuations than the 2W model.
The two-way coupled simulation with each model produces PDFs with $- 3$ power-law tails for the normal fluid velocity fluctuations.

As reported in Ref. \cite{NC-ring}, whether to use the 2W or S2W model is a very important and hot topic, when considering the two-way coupled dynamics in superfluid $^4$He. In this study, the advantages and limitations of the two mutual friction models are provided not only in elementary processes, i.e., vortex ring propagation and reconnection but also in complicated quantum turbulence. These findings support selecting the appropriate mutual friction model in each flow. At this moment, it is highlighted that the S2W model is suitable to represent the elementary processes, whereas for the quantum turbulence the 2W and S2W models have pros and cons, respectively. At the coarse-graining limit, the S2W model should approach the 2W model. The modeling of mutual friction to satisfy such a coarse-graining limit would be required in the future.

\begin{acknowledgments}
H. K. acknowledges the support from JSPS KAKENHI (Grant Number JP22H01403).
S. Y. acknowledges the support from JSPS KAKENHI (Grant Number JP23K13063).
M. T. was supported by JSPS KAKENHI (Grant Numbers JP22H05139 and JP23K03305).
\end{acknowledgments}

\newpage

\appendix*
\section{Summary of the 2W and S2W models}
\label{app}

We summarize and compare the parameters, forces, and equations of motion of the 2W and S2W models in Table \ref{table-model}.

\begin{table}[h]
\caption{Comparison of the 2W model and S2W model}
\label{table-model}
\begin{tabular}{ll}\hline\hline
2W model & S2W model \\ \hline
$\alpha = \frac{\rho_s \kappa \gamma_0}{\gamma_0^2+(\rho_s \kappa - \gamma_0')^2}$ & $\beta = \frac{\rho_s \kappa D}{D^2+(\rho_s \kappa + \rho_n \kappa)^2}$ \\
$\alpha' = \frac{\gamma_0^2-(\rho_s \kappa - \gamma_0')\gamma_0'}{\gamma_0^2+(\rho_s \kappa - \gamma_0')^2}$ & $\beta' = \frac{D^2+(\rho_s \kappa + \rho_n \kappa)\rho_n \kappa}{D^2+(\rho_s \kappa + \rho_n \kappa)^2}$ \\
$\gamma_0 = \frac{\rho_s \kappa \alpha}{\alpha^2+(1-\alpha')^2}$ & $D = \frac{\rho_s \kappa \beta}{\beta^2+(1-\beta')^2}$ \\
$\gamma_0' = \frac{\rho_s \kappa \left[ \alpha^2-\alpha'(1-\alpha')\right]}{\alpha^2+(1-\alpha')^2}$ & $\rho_n \kappa = \frac{\rho_s \kappa \left[ -\beta^2+\beta'(1-\beta)\right]}{\beta^2+(1-\beta')^2}$ \\
$\bm{f}_M=\rho_s \kappa \bm{s}'\times(\dot{\bm{s}}-\bm{v}_s)$ & $\bm{f}_M=\rho_s \kappa \bm{s}'\times(\dot{\bm{s}}-\bm{v}_s)$ \\
$\bm{f}_{D}=-\gamma_0 \bm{s}'\times \left[ \bm{s}'\times(\bm{v}_n-\dot{\bm{s}}) \right]+\gamma_0' \bm{s}'\times(\bm{v}_n-\dot{\bm{s}})$ & $\bm{f}_D=-D \bm{s}'\times \left[ \bm{s}'\times(\bm{v}_n-\dot{\bm{s}}) \right]$ \\
 & $\bm{f}_I=-\rho_n \kappa \bm{s}'\times(\bm{v}_n-\dot{\bm{s}})$ \\
$\bm{f}_{M}+\bm{f}_D=\bm{0}$ & $\bm{f}_{M}+\bm{f}_D+\bm{f}_I=\bm{0}$ \\
$\bm{f}_{sn}=\bm{f}_D$ & $\bm{f}_{sn}=\bm{f}_D+\bm{f}_I$ \\
$\bm{f}_{sn}=-\alpha \rho_s \kappa \bm{s}'\times \left( \bm{s}'\times \bm{v}_{ns} \right)-\alpha' \rho_s \kappa \bm{s}'\times \bm{v}_{ns}$ & $\bm{f}_{sn}=-\beta \rho_s \kappa \bm{s}'\times \left( \bm{s}'\times \bm{v}_{ns} \right)-\beta' \rho_s \kappa \bm{s}'\times \bm{v}_{ns}$ \\
$\dot{\bm{s}}=\bm{v}_s+\alpha s'\times\bm{v}_{ns}-\alpha'\bm{s}'\times(\bm{s}'\times\bm{v}_{ns})$ & $\dot{\bm{s}}=\bm{v}_{s\perp}+\beta \bm{s}'\times \bm{v}_{ns}-\beta' \bm{s}'\times \left( \bm{s}'\times \bm{v}_{ns} \right)$ \\ \hline\hline
\end{tabular}
\end{table}


\end{document}